# Tri-comb spectroscopy

**Bachana Lomsadze, Brad C. Smith, and Steven T. Cundiff**[*]

Department of Physics, University of Michigan, Ann Arbor, Michigan 48109, USA

*Correspondence to: cundiff@umich.edu

**Multi-dimensional coherent spectroscopy (MDCS)[1,2] is a powerful method for optical spectroscopy that has become an important tool for studying ultrafast dynamics in a wide range of systems. It is an optical analog of multidimensional nuclear magnetic resonance spectroscopy that enables the measurement of homogeneous linewidths in inhomogeneously broadened systems, many-body interactions, and coupling between excited resonances, all of which are not simultaneously accessible by any other linear or non-linear method. Current implementations of MDCS require a bulky apparatus and suffer from resolution and acquisition speed limitations that constrain their applications outside the laboratory[3-5]. Here we propose and demonstrate an approach to nonlinear coherent spectroscopy that utilizes three frequency combs with slightly different repetition rates. Unlike traditional nonlinear methods, tri-comb spectroscopy uses only a single photodetector and no mechanical moving elements to enable faster acquisition times, while also providing comb resolution. As a proof of concept, a multidimensional coherent spectrum with comb cross-diagonal resolution is generated using only 365 ms of data. These improvements make multidimensional coherent spectroscopy relevant for systems with narrow resonances (especially cold atomic and molecular systems). In addition the method has the potential to be field deployable for chemical sensing applications.**

Rapid, high precision, and sensitive spectroscopic measurements of materials are desirable both in the laboratory and for field applications such as chemical sensing and atmospheric monitoring. The development of frequency comb technology led to a method known as Dual-Comb Spectroscopy (DCS) that emerged as a revolutionary approach to optical spectroscopy [6,7].

DCS is similar to Fourier Transform Infrared (FTIR) spectroscopy but uses no moving optical elements and allows the measurement of broad, high resolution linear absorption spectra rapidly and with high sensitivity. In addition, DCS is becoming compact and field deployable by leveraging the development of micro-resonator combs [8-10]. Because of these qualities, DCS is now used both for fundamental science and for many practical applications outside the laboratory [11-16]. However DCS is a linear optical method and unable to measure the natural linewidth of inhomogeneously broadened systems, probe ultrafast dynamics or identify if resonances are coupled, all of which are critical information especially for chemical sensing applications [17].

To access this information, non-linear spectroscopic methods are required. Over the years, several nonlinear methods have been developed such as pump-probe spectroscopy, photon echo spectroscopy, stimulated Raman spectroscopy, and multidimensional coherent spectroscopy (MDCS) [18-20]. Each method has its own merits, but MDCS is considered to be the most advanced and powerful nonlinear method as it decouples homogenous and inhomogeneous linewidths, can probe extremely weak many-body interactions and give insight about the interactions between resonances[2]. However the typical experimental apparatus for MDCS incorporates bulky spectrometers, complex phase cycling schemes (to suppress linear background signals) and/or mechanical moving elements that limit the resolution and/or acquisition speed[3-5,21]. These considerations are especially problematic when measuring systems with long dephasing rates (few to tens of nanoseconds). In addition a bulky and delicate apparatus is prohibitive for use for applications outside the laboratories.

Recently frequency combs have been applied to pump-probe and multidimensional coherent spectroscopy [17,22-24] providing significant improvements in resolution and acquisition time. However, in these experiments, the time delay between the excitation pulses was still controlled using a mechanical stage while dual-comb methods were used to characterize the signal.

To fully leverage the advantages provided by frequency combs, we propose and demonstrate an approach to nonlinear spectroscopy using three frequency combs. This approach, which we call Tri-Comb Spectroscopy (TCS), enables the measurement of multidimensional coherent spectra rapidly (under 1 sec), with comb resolution and involves no mechanical moving elements.

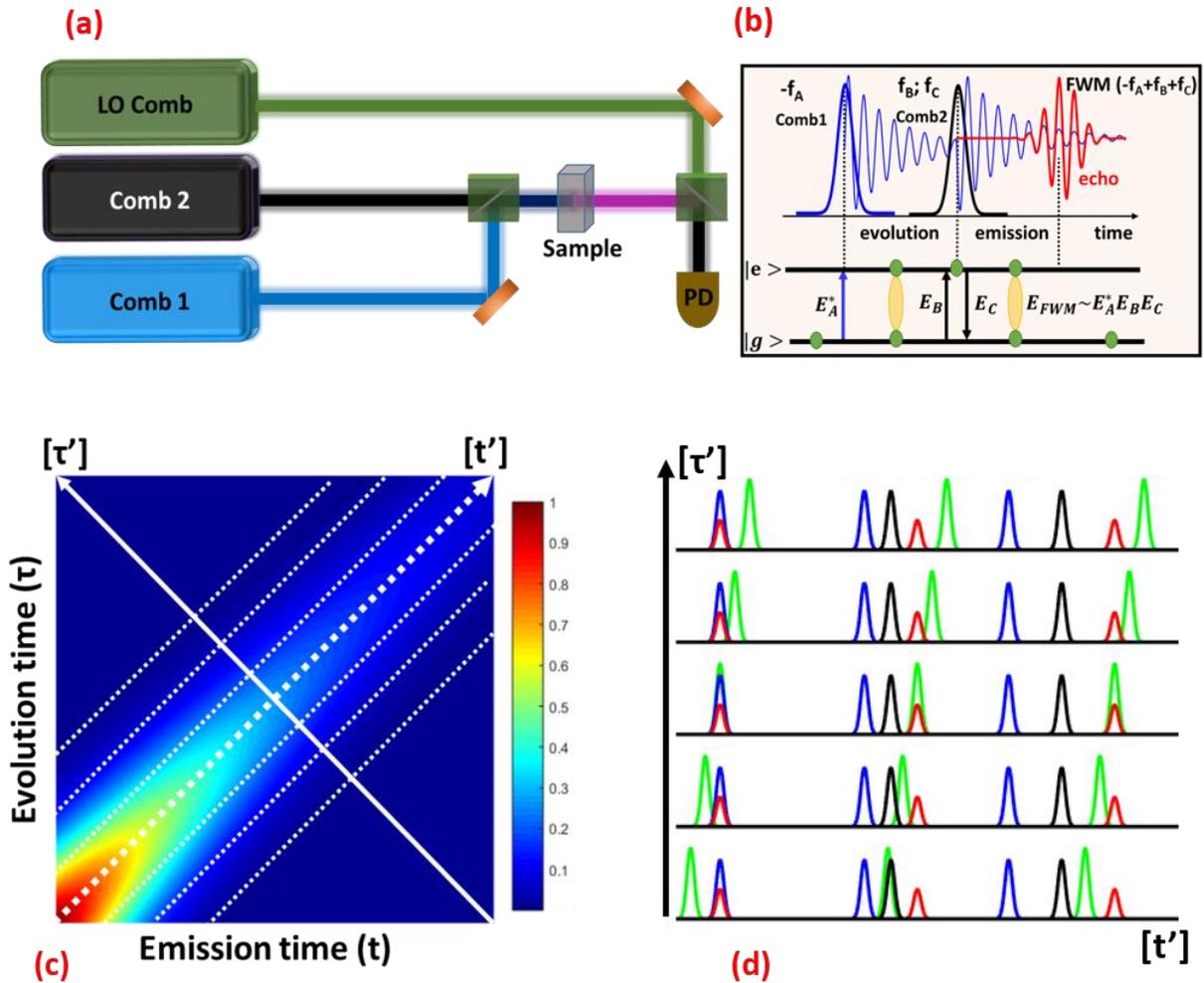

**Figure 1| Experimental setup**. (a) Experimental setup for tri-comb spectroscopy. (b) The generation of a FWM signal. The first pulse from comb 1 (blue) which is a complex phase-conjugated pulse ($E_A^*$) creates a coherence between the ground state and an excited state (evolution for a single resonance is shown in light blue); the second pulse from comb 2 (black) converts this coherence into a population of the excited state and then converts this population into a third order coherence that radiates a FWM signal which for an inhomogeneously broadened system is photon echo (red). The FWM signal is then heterodyned with the local oscillator comb. (c) Cartoon showing the magnitude of a FWM signal as a function of evolution ($\tau$) and emission (t) times. (d) In an "echo-scan", [t'] and [$\tau$'] correspond to axes of the rotated coordinate system. The cartoon shows LO pulses (green) sweeping through the linear and FWM pulses for different LO comb repetition phase.

The experimental scheme for tri-comb spectroscopy is shown in Fig. 1 (a). We used three home-built Kerr-lens mode-locked Ti:Sapphire lasers (optical spectrum centered at 800 nm) and locked the phases of the repetition frequencies (~93.5 MHz) to a four channel direct digital synthesizer (DDS). Path length and offset frequency fluctuations for each comb were measured and corrected [25] (please see the supplementary material for details). Pulses from comb 1 and comb 2 (with slightly different repetition rates) generated a four-wave-mixing (FWM) signal in the photon echo [19] excitation scheme (Fig. 1 (b)). The emitted FWM signal was then sampled and spectrally isolated (from the linear signals) in the radio frequency (RF) domain after interfering with the local oscillator (LO) comb (with a repetition frequency different from comb 1 and comb 2) on a photodetector (Suppl. Material). Fig. 1 (c) is a cartoon showing the magnitude of a FWM signal as a function of the evolution (t) and emission ($\tau$) times corresponding to the photon echo emitted by an inhomogeneously broadened system. The signal is diagonally elongated[26] (which is due to the emission of a photon echo) and is non-zero only in a limited region, hence to optimize the acquisition speed we acquired data only in this region. We refer to this data acquisition procedure as an "echo-scan". Experimentally the echo-scan was implemented by setting the relative repetition frequency between Comb 2 and LO comb ($f_{rep\_2} - f_{rep\_LO}$=274 Hz) to be exactly equal to the relative repetition frequency between Comb 1 and Comb 2 ($f_{rep\_1} - f_{rep\_2}$=274 Hz). The signs of these differences determines the time ordering of the pulses at a given point in time. This corresponds to sampling the FWM signal along the echo or along a diagonal line ([t'] axis) in Fig. 1 (c). In order to sample points off the diagonal we stepped the phase of the DDS (0.67 radian/step) serving as the reference for the LO comb (which time shifts the LO pulses by 20 ps/step but without using a delay line), measuring the FWM signal along the lines parallel to the diagonal (dashed white lines shown in Fig. 1 (c)). The concept of the "echo-scan" is shown in the time-domain in

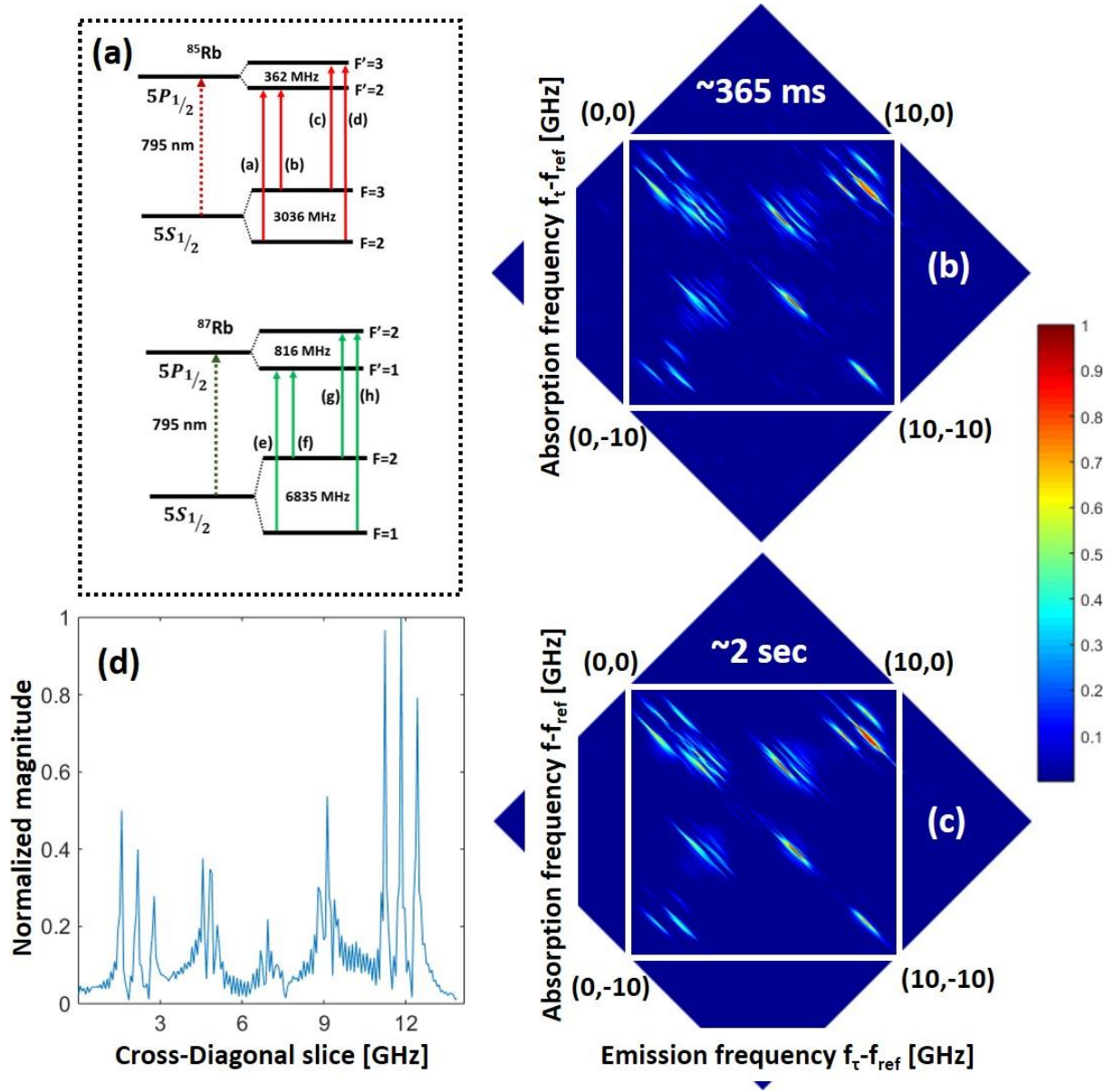

**Figure 2| |Results.** (a) Energy level diagram of $^{87}$Rb and $^{85}$Rb D1 hyperfine lines. (b) and (c) two-dimensional spectra with 365 ms and 2 second duration data records, respectively. (d) cross-diagonal slice of (c) along the (0,-10) to (10,0) GHz line. The negative values on the absorption axis reflect the negative phase evolution during the evolution period. $f_{ref}$= 377.10325 THz

Fig. 1 (d). After digitizing the signal, a multidimensional coherent spectrum was constructed by calculating a two-dimensional Fourier transform with respect to t' and τ'.

To demonstrate the resolution and acquisition speed improvements achievable with TCS we repeated the measurement shown in Figure 3. B of Lomsadze and Cundiff [17], which was taken in HVVH configuration (H-horizontal, V-Vertical polarization). We applied TCS to a vapor of Doppler broadened Rubidium atoms at 100 degree C containing both naturally occurring $^{87}$Rb and $^{85}$Rb isotopes. The energy level diagrams of Rb $D_1$ hyperfine levels for both isotopes are shown in Fig. 2 (a). In this experiment, comb 1 and comb 2 were horizontally and vertically polarized respectively and they were filtered using an optical bandpass filter centered at 794 nm (3 nm FWHM) to excite only the $D_1$ resonances. The power per beam was 2 mW and 4 mW for comb 1 and comb 2.

The results are shown in Fig. 2 (b). The two-dimensional spectrum is tilted by 45 degrees to show the spectrum in the ($f_t$ and $f_\tau$) coordinate system. The diagonal peaks (along the (0,0) to (10,-10) GHz line) correspond to absorption and emission at the same hyperfine resonances (a-h in Fig.2 a) and the cross-diagonal peaks correspond to all possible couplings between the resonances for each isotope. The peaks are diagonally elongated due to Doppler Broadening whereas along the cross-diagonal direction the widths reflect the homogenous linewidths whose resolution is now only limited by the combs' free spectral range ($f_{rep}$). The comb resolution in the cross-diagonal direction is demonstrated in Fig 2. (d) (sharp peaks correspond to single spectral points) that represents a cross-diagonal slice along the (0,-10) to (10, 0) GHz line. Comparing the spectrum in Fig. 2 (b) to reference [17] shows that we have reproduced the same results but with better resolution (factor of ~ 4 improvement) and with the "effective" acquisition time of 365 ms (almost 600 times improvement). Due to the Doppler broadening we can limit the width of the scan in the τ' direction to 1.4 ns, but for a system that is not Doppler broadened the full τ' range is required. We also note that due to overhead in data storage and instrument control, the actual acquisition time was longer than the effective time, however the latency can be easily removed with technical improvements. We acquired multiple bursts (multiple sweeps along each diagonal line in Fig.1 c) for each LO-DDS phase, but in Fig 2.B we plot the 2D spectrum that is constructed by a single burst for each LO-DDS phase, which is the minimum acquisition length. Figure 2 (c) shows the same spectrum but with the "effective" acquisition time of 2 secs (averaged over 5 bursts) that clearly demonstrates an improvement in the signal to noise ratio. However all peaks can be identified in the 365 ms data.

In conclusion we have demonstrated an approach to nonlinear optical spectroscopy using three frequency combs, tri-comb spectroscopy. We show that TCS provides a comb-resolution multidimensional coherent spectrum in under one second, which is not achievable by any other currently available methods. These improvements make multidimensional coherent spectroscopy relevant for atomic systems (both cold and Doppler broadened) and molecular ro-vibrational spectroscopy. In addition, TCS contains no delay lines, mechanical or otherwise, and can become field deployable for chemical sensing application if implemented using compact fiber lasers [27,28], mode-locked diode lasers or micro-combs [8-10,29]. TCS could also be used for real-time nonlinear hyperspectral imaging.

**Acknowledgment:** The research is based on work supported by the Office of the Director of National Intelligence (ODNI), Intelligence Advanced Research Projects Activity (IARPA), via contract 2018-18020600001. The views and conclusions contained herein are those of the authors and should not be interpreted as necessarily representing the official policies or endorsements, either expressed or implied, of the ODNI, IARPA, or the U.S. government. The U.S. government is authorized to reproduce and distribute reprints for governmental purposes notwithstanding any copyright annotation thereon. The data shown in the plots within this paper and other findings of this study are available from the corresponding author upon reasonable request.

B.C. S. acknowledges support by the National Science Foundation through a Graduate Research Fellowship (1256260);

# Supplementary Materials for

## Tri-comb spectroscopy

**Bachana Lomsadze, Brad. C. Smith and Steven T. Cundiff**

correspondence to: **cundiff@umich.edu**

**This PDF file includes:**

    Materials and Methods
    SupplementaryText
    Fig. S1
    Captions for S1
    Fig. S2
    Captions for S2

**Materials and Methods**

Method

Figure S1 shows a schematic diagram for tri-comb spectroscopy. We used three home-built Kerr-lens mode-locked Ti:Sapphire lasers centered at 800 nm. The repetition frequencies for comb 1, comb 2, and the LO comb ($f_{rep\_1} = f_{rep\_2} + 274$ Hz, $f_{rep\_2} = 93.5$ MHz and $f_{rep\_LO} = f_{rep\_2} - 274$ Hz) were phase locked to a four-channel direct digital synthesizer (DDS) but the comb offset frequencies were not actively stabilized. The average output power and pulse duration for comb 1, comb 2, and the LO comb were 250mW, 12 fs, 150 mW, 12 fs, and 150 mW, 10 fs respectively.

The output of comb 1 and comb 2 were combined on a polarizing beam splitter (PBS 1) and spectrally filtered (using an optical bandpass line filter - 3 nm full-width at half-maximum) to excite only the D1 lines of the Rb atoms. Combs 1 and 2 (with average powers after the filter of 2.0 and 4.0 mW respectively) were then focused to a 5 µm spot on the sample (loaded in a 0.5 mm thick cell and heated to $100^0$C). The four-wave-mixing (FWM) signals emitted by the sample along with the incident beams were combined with the LO comb on PBS 2 and interfered on Det 1. Half wave plates were adjusted such that some fraction of the light from each beam (comb 1, comb 2 and LO comb) was sent to PBS 3 to measure the optical phase fluctuations using the Continues Wave (CW) laser.

Material

Rb sample (99.6 % purity) was purchased at *Sigma-Aldrich* (part # 276332-1G).

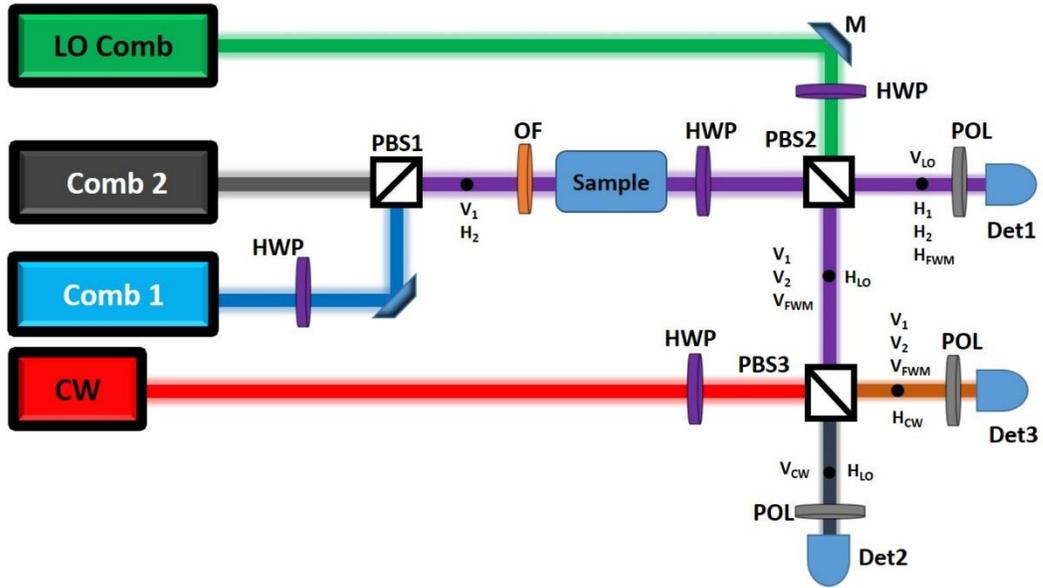

**Fig. S1** Experimental setup: LO- local oscillator, CW-continuous wave laser, HWP-Half Wave Plate, PBS-polarizing Beam Splitter, M-Mirror, POL-Polarizer, OF-Optical Filter, Det-photodetector. H-Horizontal and V-Vertical Linear polarization states of the beams.

**Supplementary Text**

**Tri-comb spectroscopy in the optical and radio frequency domain**

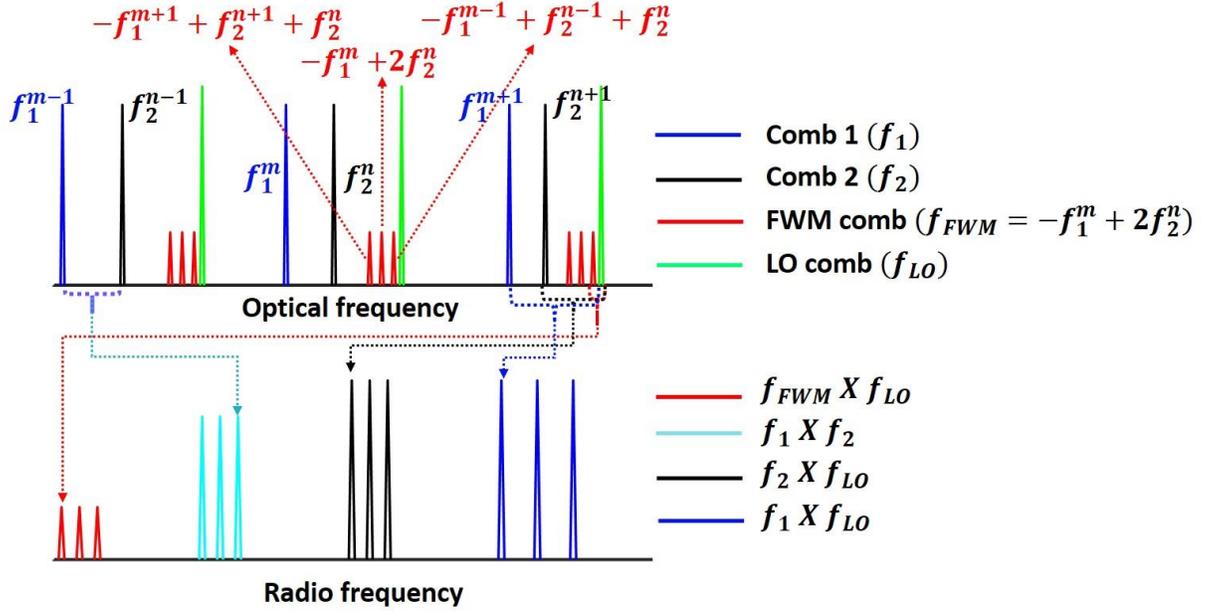

**Fig. S2** Tri-comb spectroscopy in the optical and radio frequency domains.

Figure S2 shows a schematic diagram for tri-comb spectroscopy in the optical and radio frequency domains. In the top figure (optical frequency domain) the blue, black, and green lines correspond to comb lines for Comb 1, Comb 2, and the LO comb respectively while the red lines correspond to FWM comb lines. Each FWM line has contributions from multiple combinations of $m$ and $n$. The space between adjacent lines in a given group of FWM comb lines (for example the middle group of three red lines in Fig. S2) is the difference between the repetition frequencies of the excitation combs ($f_{rep\_1} - f_{rep\_2} = 274$ Hz) whereas the repetition frequency for FWM comb is equal to the repetition frequency of the LO comb ($-f_{rep\_1} + 2 f_{rep\_2} = f_{rep\_LO}$).

Bottom figure shows the separation of linear and FWM signals in the radio frequency domain. The red, black, and blue lines are the result of heterodyning FWM, comb 2, and comb 1 teeth with the LO comb respectively whereas the cyan lines correspond to beat signals between comb 1 and comb 2.

Phase correction

We used a CW external cavity diode laser to track the phase fluctuations due to path length, offset frequency, and residual repetition frequency fluctuations for each comb. Figure S1 shows the experimental setup. The CW laser, tuned near 794 nm, was split into two parts using a HWP and PBS 3. One part was used to measure the optical phase fluctuations of LO comb on Det 2. The second part was used to measure the optical phase fluctuations of Comb 1 and Comb 2 on Det 3.

The electrical signals from Det 2 and Det 3 were digitized with a data acquisition board (100 MHz sampling rate) and used to generate a correction signal using a scheme that is similar to the one in reference [1]. The electrical signal from Det 1 containing the FWM and linear signals was amplified and digitized on the second channel of the data acquisition board. The digitized FWM signal was then mixed with the correction signal and the linear signals were used as triggers to monitor the evolution (starting point) of t and tau (Fig. 1C main manuscript). Data analysis was performed using Matlab but the phase correction process could be performed in real time using a field-programmable gate array (FPGA).

1   Lomsadze, B. & Cundiff, S. T. Frequency comb-based four-wave-mixing spectroscopy. *Opt. Lett.* **42**, 2346-2349, doi:10.1364/OL.42.002346 (2017).